\documentclass[preprint,preprintnumbers,amsmath,amssymb]{revtex4}
\usepackage[dvipdfm]{graphicx}
\begin{document}
\preprint{HUPD1203}
\def\tbr{\textcolor{red}}
\def\tcr{\textcolor{red}}
\def\ov{\overline}
\def\nn{\nonumber}
\def\f{\frac}
\def\beq{\begin{equation}}
\def\eeq{\end{equation}}
\def\bea{\begin{eqnarray}}
\def\eea{\end{eqnarray}}
\def\bsub{\begin{subequations}}
\def\esub{\end{subequations}}
\def\dc{\stackrel{\leftrightarrow}{\partial}}
\def\ynu{y_{\nu}}
\def\ydu{y_{\triangle}}
\def\ynut{{y_{\nu}}^T}
\def\ynuv{y_{\nu}\frac{v}{\sqrt{2}}}
\def\ynuvt{{\ynut}\frac{v}{\sqrt{2}}}
\def\d{\partial}
\title{Contribution from the interaction Hamiltonian  to \\ the expectation value of particle number  with \\ the non-equilibrium quantum field theory} 
\author{Ryuuichi Hotta$^1$,Takuya Morozumi$^1$,Hiroyuki Takata$^2$}
\affiliation{$^1$Graduate School of Science, Hiroshima University, Higashi-Hiroshima 739-8526,Japan \\
$^2$Tomsk state Pedagogical University Tomsk 634041,Russia
}
%\date{\today} 
%%%%%%    TEXT START    %%%%%%
\def\nn{\nonumber}
\def\beq{\begin{equation}}
\def\eeq{\end{equation}}
\def\bei{\begin{itemize}}
\def\eei{\end{itemize}}
\def\bea{\begin{eqnarray}}
\def\eea{\end{eqnarray}}
\def\ynu{y_{\nu}}
\def\ydu{y_{\triangle}}
\def\ynut{{y_{\nu}}^T}
\def\ynuv{y_{\nu}\frac{v}{\sqrt{2}}}
\def\ynuvt{{\ynut}\frac{v}{\sqrt{2}}}
\def\s{\partial \hspace{-.47em}/}
\def\ad{\overleftrightarrow{\partial}}
\def\ss{s \hspace{-.47em}/}
\def\pp{p \hspace{-.47em}/}
\def\bos{\boldsymbol}
%%%%%%    TEXT START    %%%%%%
\begin{abstract}
We develop the method analyzing particle number non-conserving phenomena
with non-equilibrium quantum fieldtheory . 
 In this study, we consider a CP violating model with interaction Hamiltonian that breaks particle number conservation.
 To derive the quantum Boltzmann equation for the particle number, we solve Schwinger-Dyson equation,
which are obtained from two particle irreducible closed-time-path (2PI CTP) effective action.
 In this calculation , we show the contribution from interaction Hamiltonian to the time evolution of expectation value of particle number.
\end{abstract}
\maketitle
\section{Introduction}
 $~$There are several particle number violating phenomena in the elementary particle physics such as leptogenesis. We are intereted in them when they occur in the non-equilibrium state.  
To understand  how particle number is produced, we study a simple model with particle number violating interaction. 
We treat the interaction perturbatively and show how the interaction Hamiltonian contribute to the expectation value of the particle number.
Throught out our paper, we use the 2PI CTP  formalism. The 2PI CTP  formalism is a method which leads to the Green functions for equilibrium and non-equilibrium states. The stationaly condition of 2PI effective action determines the Schwinger-Dyson equation for Green functions. The Green functions depend on the initial density matrix which represents the initial statistical state of the system. 
In the next section, we introduce our model and show the process which leads to the particle number violation. 
\section{Model}
$~$We consider the model  that creates the particle number. In figure 1, we show how particle number can be produced in our model. The top pannel in the figure corresponds to the particle number decreasing process while the bottom figure corresponds to the particle number creating process.The difference of the rate asymmetry calculated from both of the diagrams corresponds to the net particle number. 
In the figure 1, A is a coupling constant which violates the particle number, while $A_{\phi}$ corresponds to a coupling constant which conserves the particle number .
$B^{2}$ is a mass term which violates the particle number.   The Lagrangian which describes four processes in figure 1 is given by, 
\begin{equation}
\mathcal{L}=\frac{1}{2}{\partial}_{\mu}N{\partial}^{\mu}N -\frac{1}{2}m_{N}^{2}N^{2} + |{\partial}_{\mu}{\phi}|^{2} - m_{\phi}^{2}|{\phi}|^{2} + B^{2}{\phi}^{2} + {B^{*}}^{2}{{\phi}^{*}}^{2} + AN{\phi}^{2} + A^{*}N{{\phi}^{*}}^{2} + A_{\phi}N{\phi}{\phi}^{*},
\label{eq: eq1}\end{equation}
where N is a neutral scalar field and the particle number is assigned to be zero.  ${\phi}$ is a complex scalar field which particle number is assigned to be 1.
The Lagrangian contains two sources of particle number violation. One is an interaction term  $A_{\phi}$, and the other is the mass term $B^{2}$. 
A CP  violating phase of  the present model corresponds to the  phase difference between coupling constants A and B. It is given as,
\begin{equation}
{\phi}_{A} = \arg [\frac{A}{B^{2}}].
\end{equation}
\begin{figure}[bhtp]
\begin{center}
\includegraphics[width=6cm]{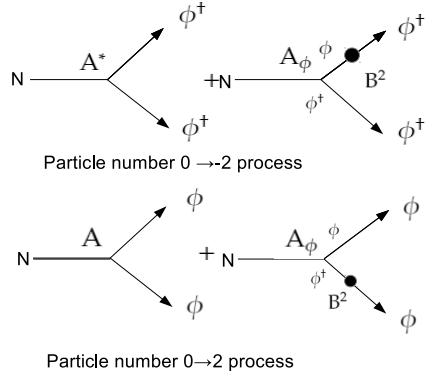}
\end{center}
\caption{Feynman diagrams which are related to the particle number violation.}
\end{figure}
 We rewrite the complex scalar field  by introducing  the two real scalar fields,  ${\phi}_{1}$ and ${\phi}_{2}$.
\begin{eqnarray}
{\phi}=\frac{{\phi}_{1}+i{\phi}_{2}}{\sqrt{2}} , \\
{\phi}^{*}=\frac{{\phi}_{1}-i{\phi}_{2}}{\sqrt{2}}.
\end{eqnarray}
Then the Lagrangian of eq. (\ref{eq: eq1}) is written as,
\begin{eqnarray}
\mathcal{L}=\frac{1}{2}{\partial}_{\mu}N{\partial}^{\mu}N -\frac{1}{2}m_{N}^{2}N^{2} + \frac{1}{2} \sum_{i=1}^{2}{\partial}_{\mu}{\phi}_{i}{\partial}^{\mu}{\phi}^{i} + \frac{1}{2} \sum_{i=1}^{2} m_{i}^{2}{\phi}_{i}^{2} +{\phi}_{i}A_{ij}{\phi}_{j}, 
\label{eq: eq5}
\end{eqnarray}
where $m_{1}$ and $m_{2}$ are masses of the real scalar fields ${\phi}_{1}$ and ${\phi}_{2}$.
They are given as ,
\begin{eqnarray}
m_{1}=(m_{\phi}^{2} - B^{2} )^{1/2},
m_{2}=(m_{\phi}^{2} + B^{2} )^{1/2}.
\end{eqnarray}
Note that the masses of  the real scalar fields are  splitted into $m_{1}$ and $m_{2}$ by the influence of $B^{2}$ term.
$A_{ij}$ denotes the interaction of N $\leftrightarrow {\phi}_{i} {\phi}_{j}$ and is given as,
\begin{eqnarray}
A_{ij}=\left[
\begin{array}{ccc}
|A|\cos [{\phi}_{A}] +\frac{A_{\phi}}{2}&-|A| \sin [{\phi}_{A}] \\
-|A| \sin [{\phi}_{A}] & |A|\cos [{\phi}_{A}] -\frac{A_{\phi}}{2} 
\end{array}
\right] .
\end{eqnarray}
In the following sections, we use the Lagrangian in eq.(\ref{eq: eq5}). 
\section{Relation between U(1) current and Green functions}
The particle number which we describe in the previous section corresponds to the U(1) charge associated with the phase transformation of the complex scalar field ${\phi}$.
In this section, we relate the U(1) current to a Keldysh Green function \cite{Keldysh:1964ud}. 
In general, there are four types of Keldysh Green functions.
Their definitions are, 
\begin{eqnarray}
&&G_{ij}^{11}(x.,y)=<T{\phi}_{i}(x){\phi}_{j}(y)> \\
&&G_{ij}^{22}(x,y)=<\tilde{T}{\phi}_{i}(x){\phi}_{j}(y)> \\
&&G_{ij}^{12}(x,y)=<{\phi}_{j}^{2}(y){\phi}_{i}^{1}(x)> \\
&&G_{ij}^{21}(x,y)=<{\phi}_{i}^{1}(x){\phi}_{j}^{2}(y)>,
\end{eqnarray}
where the matrix elements implies the trace with the initial density matrix $\rho$ as, $<A>=Tr[A{\rho}]$.
Using Keldysh Green function, the U(1) current can be written as,
\begin{equation}
<j_{\mu}(X)>=[ {{\partial}_{x}}_{\mu}G_{12}^{12}(x,y) - {{\partial}_{y}}_{\mu}G_{12}^{12}(x,y)]|_{x=y=X}.
\end{equation}
$~$In this way, we can compute the U(1) current with the Keldysh Green functions.
\section{Keldysh Green functions from 2PI effective action}
$~$In this section,we derive the Schwinger-Dyson equation for the interacting field theory in non-equilibrium enviroment.
There is the 2PI CTP formalism which introdued by E.Calzetta and B.L.Hu \cite{Calzetta:1986cq} to treat the non-equilibrium state. This method applies to our model.  The 2PI effective action of our model is given as  
\begin{eqnarray}
{\Gamma}=&&\frac{1}{2}  \log \det (G_{N})^{-1} 
+ \frac{1}{2} \textstyle\sum\limits_{i,j=1}^{2} [\log \det(G_{ij})^{-1}] 
+ \frac{1}{2} \frac{{\partial}^{2}S}{ {\partial}_{N}^{a}(x){\partial}N^{b}(y)}G_{N}^{ab}(x,y)
+ \frac{1}{2} \textstyle\sum\limits_{i,j=1}^{2} \frac{{\partial}^{2}S}{{\partial} {\phi}_{i}^{a}(x){\partial}{\phi}_{j}^{b}(y)}G_{ij}^{ab}(x,y) \nn \\
&& + {\Gamma}_{2} + const,
\label{eq: eq14}\end{eqnarray}
where the action S can be written,
\begin{equation}
S[{\phi}^{a}]=\int d^{4}x \frac{1}{2}c_{ab}{\partial}_{\mu}N^{a}{\partial}^{\mu}N^{b}- \frac{1}{2}c_{ab}N^{a}N^{b} + \frac{1}{2}c_{ab}{\partial}_{\mu}{\phi}_{i}^{a}{\partial}^{\mu}{\phi}_{j}^{b} -\frac{1}{2}c_{ab}(m_{1}^{2}{\phi}_{1}^{a}{\phi}_{1}^{b} + m_{2}^{2}{\phi}_{2}^{a}{\phi}_{2}^{b} ),
\end{equation}
where $c_{11}=1$,$c_{22}=-1$ and the other components are 0.
\begin{figure}[bhtp]
\begin{center}
\includegraphics[width=4cm]{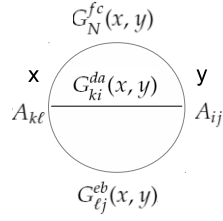}
\end{center}
\caption{2PI diagram included in ${\Gamma}_{2}$ of the 2PI effective action of eq.(\ref{eq: eq14})}
\end{figure}
By taking variation of the 2PI effective action ${\Gamma}$, we obtain the Schwinger-Dyson equation of Keldysh Green functions.
They are written as,
\begin{eqnarray}
i{\delta}^{\beta \gamma}{\delta}_{mn}{\delta}(x-y)=&&-c_{\alpha \beta} \Big{(} {\partial}_{x_{\mu}}^{2} + m_{m}^{2} \Big{)} G_{mn}^{\alpha \gamma}(x,y) 
 - \int d^{4} z {\Sigma}_{m\ell}^{\beta \alpha}(x,z)G_{\ell n}^{\alpha \gamma}(z,y) \nn \\
&&+ \int d^{4} z K_{m\ell}^{\beta \alpha}(x,z)G_{\ell n}^{\alpha \gamma}(z,y), \\
i{\delta}^{\delta \alpha}{\delta}_{mn}{\delta}(x-y) = &&-c_{\alpha \beta} \Big{(} {\partial}_{y_{\mu}}^{2} + m_{n}^{2} \Big{)} G_{mn}^{\delta \beta}(x,y)  - \int d^{4} z G_{m\ell}^{\delta \beta}(x,z){\Sigma}_{\ell n}^{\beta \alpha}(z,y) 
\nn \\
&&+ \int d^{4} z G_{m\ell}^{\delta \beta}(x,z)K_{\ell n}^{\beta \alpha}(z,y), 
\end{eqnarray}
where K is a function which is associated with the initial distribution and $\Sigma$ is a self-energy. At initial time, we assume that the expectaion value of U(1) current vanishes. Then we can focus on the time variation of the current due to the effect of interaction Hamiltonian.  
The divergence of the expected value of the current is, 
\begin{equation}
{\partial}_{\mu}<j^{\mu}(X^{\mu})>=-\int \frac{d^{3} k}{(2{\pi})^{3}}\Big{[}f_{12}^{12}(X^{0},X^{0},\mathbf{k})-e_{12}^{12}(X^{0},X^{0},\mathbf{k})\Big{]},
\end{equation}
where the f and e are the interaction terms in Schwiger-Dyson equation .Substituting the explict forms of them, the divergence of the current is given as,
\begin{eqnarray}
{\partial}_{\mu}<j^{\mu}(X^{\mu})>&=&-8 A_{1j}A_{2j} {\rm Im} [ \int \frac{d^{3}k}{(2{\pi})^{3}} 
\int \frac{d^{3}p}{(2{\pi})^{3}} {\int}_{0}^{X^{0}} dz^{0}  G_{jj}^{12}(z^{0},X^{0},\mathbf{p})G_{N}^{12}(z^{0},X^{0},-\mathbf{p}-\mathbf{k}) 
\nonumber \\
&\times& \Big{(}G_{22}^{12}(z^{0},X^{0},\mathbf{k})-G_{11}^{12}(z^{0},X^{0},\mathbf{k}) \Big{)} ].
\label{eq: eq19}\end{eqnarray}
When B is 0, then $G_{11}=G_{22}$. Therefore the current divergence vanishes.
When we compute the right hand side of eq. (\ref{eq: eq19}) in one-loop approximation, we can use the Keldysh Green functions which are derived from eq.(\ref{eq: eq14}) without ${\Gamma}_{2}$. After the time($z^{0}$) integration, the current divergence is written,
\begin{equation}
{\partial}_{\mu}<j^{\mu}(X^{\mu})>=- A_{1j}A_{2j} \int \int \frac{d^{3}pd^{3}k}{{\omega}_{i}(\mathbf{p}){\omega}_{j}(\mathbf{k}){\omega}_{N}(\mathbf{p}+\mathbf{k})(2{\pi})^{6}}\textstyle\sum\limits_{ij=\pm \pm}N_{ij} \times I^{ij}({\omega}_{i}(\mathbf{p}),{\omega}_{j}(\mathbf{k}),{\omega}_{N}(\mathbf{p},\mathbf{k})),
\label{eq: eq20}\end{equation}
where ${\omega}$ is a energy of each particle and  $N_{ij}$ denotes a combonation of bose distributin functions. 
\begin{eqnarray}
N_{++}=(n_{N}+1)(n_{j}+1)(n_{i}+1)-n_{N}n_{j}n_{i}, \\
N_{--}=(n_{j}+1)(n_{i}+1)n_{N}-n_{j}n_{i}(n_{N}+1), \\
N_{-+}=(n_{N}+1)n_{j}(n_{i}+1)-n_{N}(n_{j}+1)n_{i}, \\
N_{+-}=(n_{N}+1)(n_{j}+1)n_{i}-n_{N}n_{j}(n_{i}+1),
\end{eqnarray}
where n is bose distribution function of each particle.
Each combibation corresponds to the process shown in figure 3. 
I is a function which is given as, 
\begin{equation}
I({\omega}_{N}(\mathbf{k}+\mathbf{p})\pm {\omega}_{i}(\mathbf{k})\pm {\omega}_{j} (\mathbf{p}),X^{0})= \frac{\cos [\{ {\omega}_{N}(\mathbf{k}+\mathbf{p})\pm {\omega}_{i}(\mathbf{k})\pm {\omega}_{j} (\mathbf{p}) \} X^{0} ]-1}{{\omega}_{N}(\mathbf{k}+\mathbf{p})\pm {\omega}_{i}(\mathbf{k})\pm {\omega}_{j} (\mathbf{p})  }.
\end{equation}
From eq.(\ref{eq: eq20}), one can easily find non-zero current divergence can be obtained under the presence of the CP violating phase ${\phi}_{A}$. 
In eq.(\ref{eq: eq20}), the current divergence is given by the sum of the four diagrams shown in the figure 3 and their four inverse processes.
\begin{figure}[bhtp]
\begin{center}
\includegraphics[width=6cm]{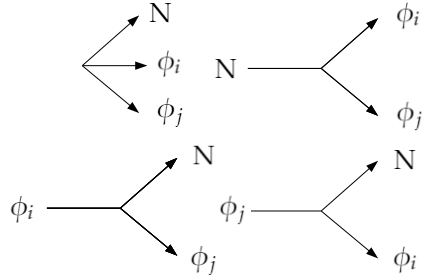}
\end{center}
\caption{Feymann diagrams which contribute to the current divergence. Their inverse processes are not shown.}
\end{figure}
\section{Summary}
$~$In this paper, we compute the current divergence of  our model by using 2PI CTP formalism.
As a result, we show the mechanism how particle number is produced. 
We find that the production rate of particle number is propotional to the CP violating phase $\phi_A$ by computing the current divergence eq. (\ref{eq: eq20}) with the Keldysh Green function. The current divergence is given as in eq.(\ref{eq: eq20}) and it is the sum of the contributions in figure 3. For numerical caluculation of the production rate, we must carry out the momentum integration of eq.(\ref{eq: eq20}) 

\end{document}